\documentstyle[12pt]{article}

\input epsf

\newcommand\figinsert[4]
{\begin{figure}[htb]
\newcommand\figsize{#3}
\epsfysize\figsize
\vskip -.5cm
\centerline{\epsffile{#4}}
\vskip -.0cm
\caption{\leftskip 1pc \rightskip 1pc
\baselineskip=10pt plus 2pt minus 0pt {{#2}}}
\label{fig #1}
\vskip -.3cm\end{figure}}

\font\grande=cmr10 scaled \magstep4
\font\medio=cmr10 scaled \magstep2

\def\laq{\raise 0.4 ex \hbox{$<$}\kern -0.8 em\lower 0.62 ex\hbox{$\sim$}}
\def\gaq{\raise 0.4 ex \hbox{$>$}\kern -0.7 em\lower 0.62 ex\hbox{$\sim$}}

\def\laq{\raise 0.4ex\hbox{$<$}\kern -0.8em\lower 0.62
ex\hbox{$\sim$}}
\def\gaq{\raise 0.4ex\hbox{$>$}\kern -0.7em\lower 0.62
ex\hbox{$\sim$}}

\begin{document}
\bibliographystyle {unsrt}
\newcommand{\pa}{\partial}

\titlepage
\begin{flushright}
DF/IST--6.97 \\
October 1997
\end{flushright}
\vspace{15mm}

\begin{center}
{\bf Primordial Density Fluctuations in a Dual Supergravity Cosmology}\\
\vspace{5mm}
{\grande }
\vspace{10mm}
M. C. Bento, O. Bertolami and N. J. Nunes\\
{\em Departamento de F\'{\i}sica,
  Instituto Superior T\'ecnico}\\

{\em Av. Rovisco Pais, 1096 Lisboa Codex, Portugal } \\
\end{center}

\vspace{10mm}
\centerline{\medio  Abstract}
\noindent
We analyse the spectrum of energy density fluctuations of a dual supergravity
model where  the dilaton and the moduli
are stabilized and sucessful inflation is achieved inside domain walls
that separate different vacua of the theory.  Constraints on  the
parameters of the superpotential are derived from
the amplitude of the primordial energy density fluctuations as inferred
from COBE and it is shown that the scale dependence of the tensor
perturbations nearly vanishes.

\vfill
\newpage

\setcounter{equation}{0}
\setcounter{page}{2}

Measurements of the cosmic microwave background anisotropy can be used to
test in detail the
inflationary hypothesis within the next decade. Hence, it is important to
study field-theoretic inflation in a realistic context. It is thought that
physics below
the Planck scale is described by a unified theory of all the fundamental interactions
incorporating supersymmetry. The local version of supersymmetry -- supergravity --
includes gravity and is, therefore, particularly relevant to build inflationary
models in the supergravity context. A generic problem for a wide variety 
of supergravity models is that  the  
effective potential for the  would-be inflaton field, $\phi$,  is too steep,
 growing  as exp${(\frac{\phi^2}{M^2})}$
for large $\phi$, to allow a sufficiently long
period of inflation to occur.  In string-motivated supergravity models where 
supersymmetry is broken via gaugino condensation, there are specific difficulties 
since, besides a steep dilaton potential preventing the use of this field as the 
inflaton, it is also difficult to implement the alternative scenario where 
the dilaton is stabilized and  inflation is driven by other fields 
\cite{brustein}, leading to a runaway problem. Furthermore, at least up to order
$\alpha^{\prime 3}$, the
structure of the higher order curvature terms does not allow, in the
presence of the dilaton, for stable de Sitter type solutions
\cite{bento1}. However, there are some cases where a successful
inflationary scenario can be developed as
in the so-called dilaton-driven kinetic inflation or Pre-Big-Bang
model \cite{veneziano} \footnote{Although this model seems to suffer
from a graceful exit problem \cite{brustein2}, which cannot be
solved with the help of higher-curvature terms \cite{kaloper}, there are
indications that a quantum cosmological approach may resolve this difficulty 
\cite{bento2}.} or assuming that the dilaton has been stabilized 
and that the inflaton is a gauge singlet field other than the dilaton 
\cite{ross,macorra,yanagida} .
A successful  scenario can  also be achieved  using  hybrid inflationary models 
\cite{linde}, where there are  
two stages of inflation driven by  the inflaton and a GUT Higgs field.
In Ref. \cite{bento3},  a proposal based on a dual superstring model
was made that circumvents the abovementioned difficulties via  topological
inflation,  realizing in a concrete fashion  a generic scenario first proposed in 
\cite{banks} to resolve the so-called cosmological moduli problem \cite{polonyi}.
In this letter, we analyse other aspects of that model, namely the
spectrum of density fluctuations and the reheating temperature.

In Ref. \cite{bento3} it is shown that, once the requirements of S and T-duality 
invariance are imposed, inflation can be achieved via the imaginary part of S since the
conditions for successful inflation are satisfied by domain walls that
separate degenerate minima  (notice, however, that T-duality alone is not compatible
with topological inflation).  
This model avoids the dilaton runaway problem since the S and T-dual 
invariant potential has minima for finite, periodic  values of the moduli. 
Moreover, topological inflation solves the problem of initial 
conditions for the onset of inflation as, due to the fact 
that inflation is driven by a topological defect, the field sits necessarily 
at the top of the potential.
 
In $N=1$ supergravity, S-duality was conjectured \cite{font} in analogy with T-duality, 
a well-established symmetry of string compactification \cite{alvarez}. Indeed, the
target space modular
invariance of string effective actions contains a duality transformation as well
as discrete shifts of the axionic background. The conjecture  is that there would
be a further modular invariance symmetry in string theory, where the modular group
now acts on the complex scalar field $S=\phi + i \chi$, where $\phi$ is the
dilaton and $\chi$ is a pseudoscalar axion field. Notice that this symmetry,
which includes S-duality, under which the dilaton gets inverted, strongly constrains
the theory since it relates the weak and strong coupling regimes.

Imposing S and T-duality on the
Lagrangian for $N=1$ supergravity theory, one  obtains the following
potential for a model with 4 moduli, $S$, $T_1$, $T_2$, and $T_3$ 
\cite{macorra} (for models where S-duality is implemented in a different way 
see \cite{lalak}):

\begin{equation}
\label{bca}
V_F=e^K \vert\eta(T_2)\eta(T_3)\eta(S)\vert^{-4}
\left(\vert P\vert^2\left[\frac{S_R^2}{4\pi^2}\vert\hat G_2(S)\vert^2 +
\frac{T_{Ri}^2}{4\pi^2}\vert\hat G_2(T_i)\vert^2 - 2\right] + F_0\right),
\end{equation}

\noindent
where $S_{R}=S+ \bar S$, $T_{Ri}=T_{i}+ \bar T_{i}$,  
$P=P(T_1,\psi)=\lambda(T_1)\Theta(\psi)$, $\psi$ denotes the
untwisted chiral fields related to the $T_1$ sector, $\lambda$ and $\Theta$
are gauge invariant functions and $F_0=P_m(K^{-1})_n^mP^n +
(K_mP(K^{-1})_{T_1}^mP^{T_1} + h.c.)$, the indices indicate, $m$ and $n$ run 
over all moduli derivatives and $i = 2,3$.  The function  
$\eta(S)=q^{1/24} \prod_n(1-q^n)$ is the 
well-known Dedekind function, $q\equiv \exp (-2\pi S)$; 
$\hat G_2(S) = G_2(S) - 2 \pi/S_R$ is the weight two Eisenstein function 
and $G_2(S)=\frac{1}{3} \pi^2 - 8 \pi^2 \sum_n \sigma_1(n) \exp(- 2\pi n S)$, where  
$\sigma_1(n)$ is the sum of the divisors of $n$, and analogously for the $T_i$ moduli. 
The K\"ahler function, $K$, and the 
superpotential, $W$, are given respectively by:

\begin{equation}
\label{bey}
K = -~lnS_{R} - 3~lnT_{R1} - 3~lnT_{R2} - 3~lnT_{R3}~,
\end{equation}

\begin{equation}
\label{bez}
W = \eta(S)^{-2}~\eta(T_2)^{-2}~\eta(T_3)^{-2}~P(T_{1},\psi)~,
\end{equation}
where the contribution of chiral matter fields was dropped.

Clearly, this potential is S (and $T_i$)-duality invariant since 
all dependence on $S$ and $T_i$ is given in terms of duality-invariant 
functions $e^K \vert\eta(S)\vert^{-4}$ and $S_R^2\vert \hat G_2(S)\vert^2$. 
The dual invariant points $<S>=1,\ e^{-\pi/6}$ and
$<T_i>=1,\ e^{-\pi/6}$ are extrema (maxima and saddle points, respectively)
and the minima of V are nearby.

In addition to  (1), we have to consider the contribution of $D$-terms associated with 
the gauge sector of the theory, namely

\begin{equation}
\label{new}
V_D = \frac{1}{2 Re f} D^2 \ ,
\end{equation}

\noindent
where $D=\hat g K^i T^j_i \Phi_j + \xi$, $\hat g$ being the gauge charge, 
$T^j_i$ are the generators of the gauge group and $\xi$ the Fayet-Illiopoulos term. 
S-duality is ensured for $f = \frac{1}{2 \pi}[\ln(j(S) - 744]$, $j(S)$ being the 
generator of modular invariant functions  \cite{lalak}. From string perturbative results 
$f=S$ and, therefore,  
S-duality implies that  \footnote{Another realization of S-duality is 
$f\rightarrow 1/f$, but this requires the presence of the so-called 
``magnetic condensate'' \cite{font,lalak}.} $f\rightarrow f$.

Assuming that  the
T-fields and the untwisted fields of the $T_1$ sector have already settled
at the minimum of the potential so that inflation takes place due to
the S-field, the  potential of eq. (1) can be then written as

\begin{equation}
\label{bef}
V_F \sim \left[ \frac{1}{S_R\vert \eta(S)\vert^4}
\left( \frac{S_R^2}{4 \pi^2}\vert \hat G_2(S) \vert^2 - a \right) \right],
\end{equation}

\noindent
where $a$ parametrises $F_0$.
Figure 1 shows the potential as a function of Re S and Im S, for
$a = 3$.

In the model of Ref. \cite{bento2},
it is further assumed that $S_R$ has settled at the minimum of the potential (at
$<S_R> \sim 2.6$). The total potential, relevant for the computation of density 
perturbations in our model is, therefore

\begin{equation}
\label{be}
V = c \left[ \frac{1}{<S_R>\vert \eta(S)\vert^4}
\left( \frac{<S_R^2>}{4 \pi^2}\vert \hat G_2(S) \vert^2 - a \right) + b \right],
\end{equation}
where
\begin{equation}
\label{bex}
c \equiv e^{<K>} \vert P \vert^2 <S_{R}>  \vert \eta(<T_2>) \eta(<T_3>)
\vert^{-4}~.
\end{equation}
and parameter $b$ has been added, representing the contribution of the ground state of 
$D$-terms, eq. (\ref{new}), and ensuring that the  potential is positive (this was an 
implicit assumption in \cite{bento3}). Notice that, at this stage, 
S-duality (and also supersymmetry) is broken since a particular non-vanishing 
vacuum state has been chosen.

In Ref. \cite{bento3}, it has been shown that the
conditions for topological inflation, a scenario first put forward by
Linde  \cite{linde} and Vilenkin \cite{vilenkin}, to occur at the core of
the domain walls separating degenerate minima of the 
potential can be met for some range of parameters.
In this scenario, inflation
takes place  as the imaginary part of the S field expands
exponentially, provided certain conditions   are satisfied at the top of the
potential. Next, we shall briefly discuss these conditions.

Along a domain wall $\chi$ ranges from one minimum in one region
of space to another minimum in another region. Somewhere between, $\chi$
must traverse the top of  the potential, $\chi_0$, and
we hence start expanding the potential about $\chi_0$

\begin{equation}
\label{ah}
V\simeq c \left[V_0 \left( 1-\alpha^2 \frac{(\chi - \chi_0)^2}{M^2}\right) + b\right],
\end{equation}

\noindent
$M$ being the natural scale of the fields in supergravity and that 
was set to one in Eqs. (1) - (5).
In flat space, the wall thickness is equal to the curvature of the
effective potential, that is $\delta^{-1} \simeq \alpha(c V_0/M_P^2)^{1/2}$. The
Hubble parameter in the interior of the wall is given by
$H\simeq (\frac{8}{3}\pi G c \hat{V_0})^{1/2}$, with $\hat{V_0}=V_{0}+b$. If
$\delta \ll H^{-1}$, gravitational effects are negligible. However, if
$\delta > H^{-1}$, the region of false vacuum near the top of the potential,
$V\simeq c \hat{V_0}$, extends over a region greater than a Hubble volume. Hence, if
the top of the potential satisfies the conditions for inflation, the interior
of the wall inflates. Demanding that $\delta > H^{-1}$, one obtains the
following condition on $\alpha$:
$\alpha^2 < \frac{8 \pi}{3} (\frac{\hat{V_0}}{V_0})$.
It turns out that this condition is more stringent than the ones
that can be derived from demanding an inflationary slow rollover regime
\cite{banks}.
However, the requirement that there are at least $N_e$ e-folds
of inflation, i.e.
$\frac{-V^{\prime\prime}}{V}\ll \frac{6 \pi}{N_e}$
leads to the most stringent constraint on $\alpha$  (for $N_e=65$)
\cite{bento2}:

\begin{equation}
\label{bb}
\alpha^2 < \frac{3\pi}{65}~~.
\end{equation}

We have computed $\alpha^2= \frac{1}{2} M^{2} \vert \frac{V^{\prime\prime}}{V}\vert$  
for different values of $a$ and $b$  and found that, in order to have a positive
potential and satisfy the condition (\ref{bb}), we must have $b~\geq~8.1~M^{4}$and

\begin{equation}
\label{bba}
 a\ \geq\ 2.5~~;
\end{equation}
furthermore, we have checked that the slow roll over conditions
$\vert\eta\vert= M^{2} \vert \frac{V^{\prime \prime}}{V}\vert \leq 1$ and
$\epsilon =\frac{1}{2} M^{2} (\frac{V^\prime}{V})^2 \leq 1$ are satisfied for
any value of $\chi$ in the relevant range, $\chi_{0} \leq \chi \leq \chi_{0} + 0.5$.

Notice that considering the non-canonical struture of
the kinetic terms of $S$ (and $T$)  dictated by
N=1 supergravity , $(S_R)^{-2} \partial_\mu S\partial^\mu S^{*}$
$((T_R)^{-2} \partial_\mu T\partial^\mu T^{*})$, does not change our results due to
modular invariance \footnote{This point was overlooked in  \cite{lyth},
where the periodic structure of the potential in the $ImS$ direction was 
approximated by a sinusoidal function, which was then transformed to 
account for the non-canonical
structure of the kinetic terms, a procedure that does not respect modular
invariance.}.

Hence, we conclude that topological inflation is possible for $a \geq\ 2.5$ 
and $b~\geq~8.1~M^{4}$,
thereby solving the initial condition problems in these models \cite{bento3}.

Of course, in order to have a complete cosmological scenario,
it is still required that primordial energy
density fluctuations are generated and a successful phase of reheating is
achieved. A constraint on the remaining parameter of the superpotential, namely $c$,
can be derived  from the spectrum of adiabatic density
fluctuations, which is given, in terms of the potential, by \cite{kolb}:

\begin{equation}
\label{ca}
\delta_H\equiv \sqrt{4 \pi} \left( \frac{\delta \rho}{\rho}\right)_H =
\frac{1}{5\sqrt{3}\pi M^3}\frac{V_{N}^{3/2}}{V_{N}^\prime}~~,
\end{equation}
where the subscript $N$ indicates that the right-hand side should
be evaluated as the comoving scale $k$ equals the Hubble radius, $k=a(t)
H(t)$, during inflation. Neglecting  higher multipoles in  the Cosmic
Microwave Background radiation observed by COBE, the best fit for the quadrupole
moment obtained from the angular power spectrum  corresponds to \cite{cobe}

\begin{equation}
\label{cb}
\delta_H\approx 2.3\times 10^{-5}
\end{equation}
with an uncertainty of about $10\%$.

On the other hand, the spectral index $n_s$ of the density fluctuations
is given in terms of the slow roll over parameters by \cite{kolb}

\begin{equation}
\label{cc}
n_s \simeq 1 - 6\ \epsilon(\chi_N) + 2\ \eta(\chi_N)
\end{equation}

In Figure 2, we show $n_s(\chi)$ for $a=3$ and $b=8.3~M^{4}$.
We see that consistency with observational bounds, i.e.
$ 0.6 \leq n_s \leq 1.2$,
requires $\chi_0 \leq \chi_N \leq \chi_0 + 0.35$ for the choice $b~\gaq~8.1~M^{4}$. 
Notice that $\delta_{H}$ depends on $c^{1/2}$ implying that,
in order  to satisfy the bound (\ref{cb}), $c$ is constrained to be in the range
$1.7 \times 10^{-11}~\leq~c~\leq~4.9 \times 10^{-11}$ (see Figure 3), where 
we have chosen the values of $\chi$ in the region  
$\chi_0 + 0.1  \leq \chi \leq \chi_0 + 0.4$ in order 
to avoid the singularity of Eq. (\ref{ca}) at the extrema of the potential. Hence
we can conclude that 

\begin{equation}
\label{ccb}
5.0 \times 10^{-10}~\leq~\vert P \vert^{2}~\leq~1.4 \times 10^{-9}~,
\end{equation}
as $\eta(T_i)=0.7$ and we have assumed that $<T_{R1}> \approx <T_{Ri}> = 2$.

As for the scale dependence of the tensor perturbations we obtain:

\begin{equation}
\label{ccx}
n_t \simeq - 2 \epsilon(\chi_N)~\laq -~5 \times 10^{-4}~,
\end{equation}
meaning that the predicted spectrum of gravitational waves is nearly scale 
invariant.
Furthermore, as the amplitude of the tensor perturbations is given by 
$\epsilon(\chi_N)^{1/2} \delta_H$ it follows this is 
about two orders of magnitude smaller than the amplitude of scalar perturbations.

Once field $\chi$ begins to
oscillate about its minimum,  the Universe undergoes a reheating phase. At minimum,
the inflaton field has a mass
$m_\chi= \sqrt{2} ~\gamma$, where $\gamma~\laq~1.4 \times 10^{-5} \alpha M$.
Since the dilaton is hidden from the other sectors of the theory, it
couples to lighter fields with strength $\sim \gamma/M$, leading to a decay
width

\begin{equation}
\label{bbb}
\Gamma_\chi \simeq \frac{m_\chi}{(2\pi)^3} \left(\frac{\gamma}{M}\right)^3,
\end{equation}
and a reheating temperature

\begin{equation}
\label{bc}
T_{RH} = \left(\frac{30}{\pi^2 g_{RH}}\right)^{1/4} \sqrt{M \Gamma_{\chi}} \simeq
\frac{2}{\pi^2}\left(\sqrt{\frac{15}{g_{RH}}}\frac{\gamma^3}{M}\right)^{1/2},
\end{equation}
where $g_{RH}$ is the number of degrees of freedom at $T_{RH}$.

A severe upper bound on $T_{RH}$ comes from the
requirement that gravitinos are not abundantly regenerated in the
post-inflationary reheating epoch. Indeed, once regenerated beyond a
certain density, stable thermal
gravitinos would dominate the energy density of the Universe or, if
they decay, have undesirable effects on nucleosynthesis and light element
photo-dissociation and lead to distortions
in the microwave background.
This implies in bounds of the type \cite{ellis}:

\begin{equation}
\label{bcaa}
T_{RH}~\laq ~(2 - 6) \times
10^9~\hbox{GeV}~~~~\hbox{for}~~~~m_{3/2} = (1 - 10)~\hbox{TeV}.
\end{equation}

For the model under consideration, for $g_{RH}\approx 150$, we get:

\begin{equation}
\label{bcab}
T_{RH}~\laq~1.4~\times 10^{-9}~M = 3.4~ \times 10^{9}~~\hbox{GeV}.
\end{equation}

However, as discussed in \cite{font}, S-duality implies that the gravitino mass is 
rather high, $O(M)$, in S-dual models without T-duality. In the model of Ref. 
\cite{macorra}, which is S and $T_i$-dual, one obtains, after
satisfying (\ref{cb}), $m_{3/2} \equiv e^{<G>/2}~M \simeq 1.3~c^{1/2}~M~
\laq~10^{-5}~M$, where $G = K + ln \vert W \vert^2$, implying
that there is actually no bound on the reheating temperature.
In models where one implements S-duality and the possibility of gaugino condensation 
\cite{lalak}, the gravitino 
mass can be much smaller and the bounds (18) may turn out to be relevant.
Finally we mention that, for our choice of parameters, the vacuum energy 
density can be estimated as $\rho_V~\laq~2~\times 10^{-10}~M^4$.

We can then summarize our results as follows. Topological or defect inflation
can be achieved in the context of $N = 1$ supergravity theories arising from $S$
and $T$ dual string models, as discussed in Ref. \cite{bento3},
and consistency with the observed amplitude of energy density
fluctuations can be obtained if the function parametrizing the untwisted fields
of the theory satisfies the condition 
$5.0 \times 10^{-10}~\leq~\vert P \vert^{2}~\leq~1.4 \times 10^{-9}$, $a~\geq~2.5$ and 
$b~\geq~8.1~M^4$.
This condition ensures that there is no gravitino problem as the
gravitino mass is rather high, $m_{3/2}~\laq~10^{-5}~M$. Furthermore, we
predict  the spectral index to be in the range $0.7 \leq n_s \leq 1.2$, as can be
seen in Figure 2, and a nearly scale independent spectrum of tensor perturbations, since
$n_t~<<~1$.


\vfill
\eject

\newpage


\figinsert{Fig1}       
{The potential as a function of Re S and Im S (for $a = 3$, $b= 8.3~M^{4}$ and
$c = 1$, see Eq. (4)).}                    
{2.0truein}{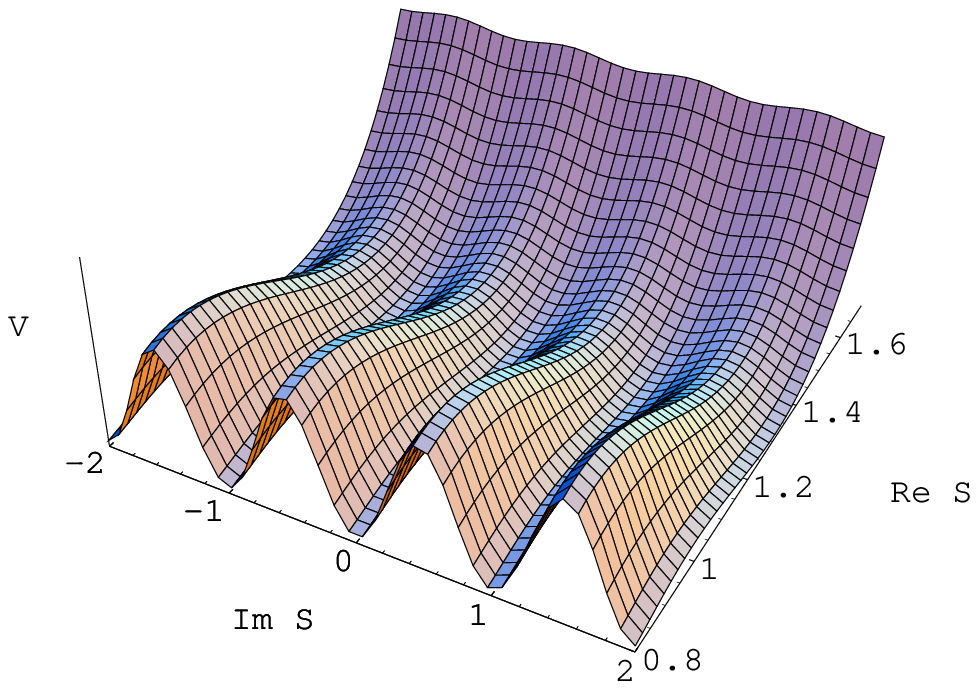}  

\vskip 3.5cm

\figinsert{Fig2}       
{The spectral index $n_s$ as a function of $\chi$ ($a = 3$ and
$b= 8.3~M^{4}$).}                    
{2.0truein}{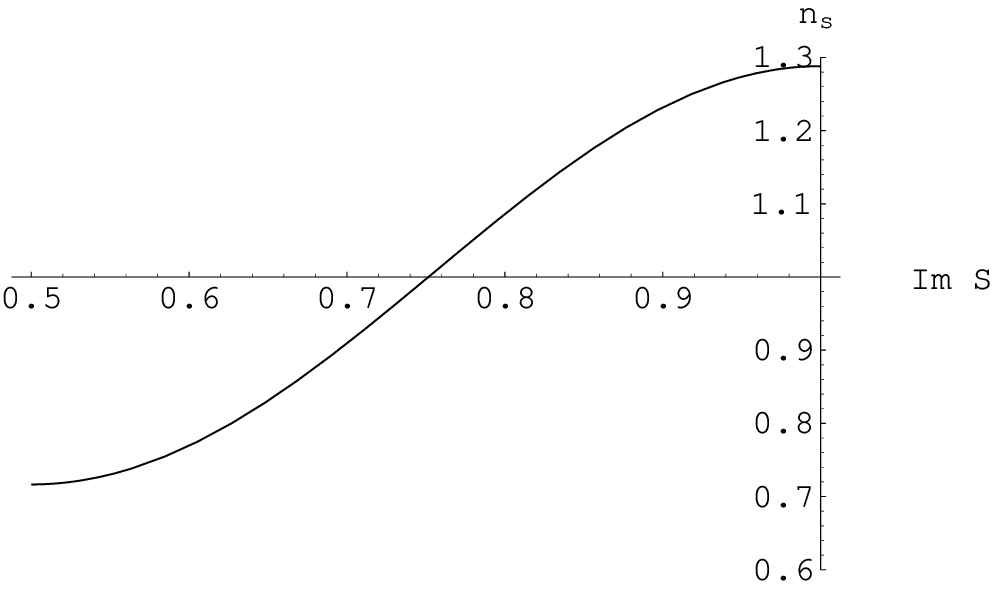}  

\newpage

\figinsert{Fig3}       
{The density fluctuation parameter $d \equiv \delta_H /c^{1/2}$
as a function of $\chi$ ($a = 3$ and $b= 8.3~M^{4}$).
}                    
{2.0truein}{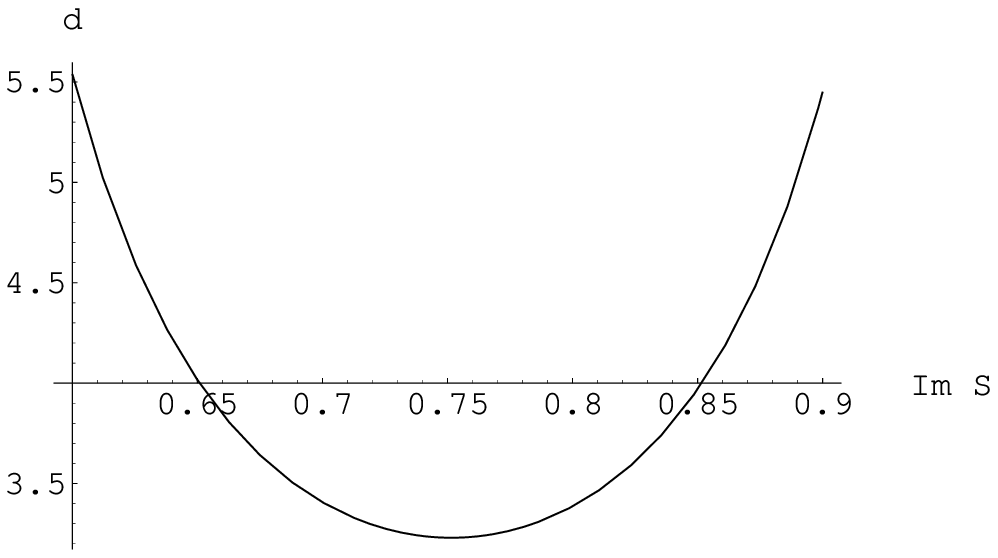}  


\end{document}